\documentclass{ws-rv975x65}
\usepackage{subfigure}     
\usepackage{ws-rv-van}     
\usepackage{harpoon}
\makeindex

\usepackage{chngcntr}
\counterwithout{chapter}{chapter}
\counterwithout{section}{section}
\counterwithout{equation}{section}
\counterwithout{figure}{section}
\counterwithout{table}{section}

\begin{document}

\chapter[Using World Scientific's Review Volume Document Style]
{Quark Mass Hierarchy and Flavor Mixing Puzzles}

\author[Zhi-zhong Xing]{Zhi-zhong Xing\footnote{E-mail: xingzz@ihep.ac.cn}}

\address{1) Institute of High Energy Physics, Chinese Academy of Sciences,
Beijing 100049; \\
2) Center for High Energy Physics, Peking University, Beijing 100080,
P.R. China.}

\begin{abstract}
The fact that quarks of the same electric charge possess a mass
hierarchy is a big puzzle in particle physics, and it must be highly
correlated with the hierarchy of quark flavor mixing. This review article
is intended to provide a brief description of some important issues
regarding quark masses, flavor mixing and CP violation. A comparison
between the salient features of quark and lepton flavor mixing
structures is also made.
\end{abstract}
\markright{Customized Running Head for Odd Page}
\body

\section{A Brief History of Flavors}

In the subatomic world the fundamental building blocks of matter are
known as ``flavors", including both quarks and leptons. Fifty years
ago, the quark model was born thanks to the seminal work done by
Murray Gell-Mann \cite{GM} and George Zweig \cite{Zweig}
independently; and it turned out to be a great milestone in the
history of particle physics. The phrase ``flavor physics" was first
coined by Harald Fritzsch and Murray Gell-Mann in 1971, when they
were trying different flavors of ice cream at one of the Baskin
Robbins stores in Pasadena. Since then quarks and leptons have been
{\it flavored}.

There are totally twelve different flavors within the Standard Model
(SM): six quarks and six leptons. Table \ref{table1} is an
incomplete list of the important discoveries in flavor physics,
which can give one a ball-park feeling of a century of developments
in particle physics. The SM contains thirteen free flavor parameters
in its electroweak sector: three charged-lepton masses, six quark
masses, three quark flavor mixing angles and one CP-violating phase.
Since the three neutrinos must be massive beyond the SM, one has to
introduce seven (or nine) extra free parameters to describe their
flavor properties: three neutrino masses, three lepton flavor mixing
angles and one (or three) CP-violating phase(s), corresponding to
their Dirac (or Majorana) nature
\footnote{In this connection we have assumed the $3\times 3$ lepton
flavor mixing matrix $U$ to be unitary for the sake of simplicity.
Whether $U$ is unitary depends on the mechanism of neutrino mass
generation.}.
So there are at least twenty flavor parameters at low
energies. Why is the number of degrees of freedom so big in the
flavor sector? What is the fundamental physics behind these
parameters? Such puzzles constitute the flavor problems in modern
particle physics.
\begin{table}[t]
\renewcommand\arraystretch{1.2}
\tbl{An incomplete list of some important discoveries in the
100-year developments of flavor (quark and lepton) physics
\cite{XZ}.
\vspace{0.2cm}}
{\begin{tabular}{l|l} \hline\hline
& Discoveries of lepton flavors, quark flavors and weak CP violation \\
\hline
  1897 & electron (Thomson \cite{Thomson})  \\
  1919 & proton (up and down quarks) (Rutherford \cite{Rutherford}) \\
  1932 & neutron (up and down quarks) (Chadwick \cite{Chadwick}) \\
  1933 & positron (Anderson \cite{Anderson}) \\
  1937 & muon (Neddermeyer and Anderson \cite{Anderson2}) \\
  1947 & Kaon (strange quark) (Rochester and Butler \cite{Rochester}) \\
  1956 & electron antineutrino (Cowan {\it et al.} \cite{Cowan}) \\
  1962 & muon neutrino (Danby {\it et al.} \cite{Danby}) \\
  1964 & CP violation in $s$-quark decays (Christenson {\it et al.}
  \cite{Christenson}) \\
  1974 & charm quark (Aubert {\it et al.} \cite{Aubert} and Abrams {\it et al.}
  \cite{Abrams}) \\
  1975 & tau (Perl {\it et al.} \cite{Perl}) \\
  1977 & bottom quark (Herb {\it et al.} \cite{Herb}) \\
  1995 & top quark (Abe {\it et al.} \cite{Abe} and Abachi {\it et al.}
  \cite{Abachi}) \\
  2001 & tau neutrino (Kodama {\it et al.} \cite{Kodama}) \\
  2001 & CP violation in $b$-quark decays (Aubert {\it et al.} \cite{Aubert2}
  and Abe {\it et al.} \cite{Abe2}) \\
  \hline
\end{tabular}}
\label{table1}
\end{table}

\section{Quark Mass Hierarchy}

Quarks are always confined inside hadrons, and hence the values of
their masses cannot be directly measured. The only way to determine
the masses of six quarks is to study their impact on hadron
properties based on QCD. Quark mass parameters in the QCD and
electroweak Lagrangians depend both on the renormalization scheme
adopted to define the theory and on the scale parameter $\mu$ ---
this dependence reflects the fact that a bare quark is surrounded by
a cloud of gluons and quark-antiquark pairs. In the limit where all
the quark masses vanish, the QCD Lagrangian possesses an $\rm
SU(3)^{}_{\rm L} \times SU(3)^{}_{\rm R}$ chiral symmetry, under
which the left- and right-handed quarks transform independently. The
scale of dynamical chiral symmetry breaking (i.e., $\Lambda^{}_\chi
\simeq 1$ GeV) can therefore be used to distinguish between light
quarks ($m^{}_q < \Lambda^{}_\chi$) and heavy quarks ($m^{}_q >
\Lambda^{}_\chi$) \cite{Gasser}.

One may make use of the chiral perturbation theory, the lattice
gauge theory and QCD sum rules to determine the {\it current} masses
of the three light quarks $u$, $d$ and $s$ \cite{PDG}. Their values
can be rescaled to $\mu = 2$ GeV in the modified minimal subtraction
($\overline{\rm MS}$) scheme, as shown in Table \ref{table2} . On
the other hand, the heavy quark effective theory, the lattice gauge
theory and QCD sum rules allow us to determine the {\it pole} masses
$M^{}_c$ and $M^{}_b$ of the charm and bottom quarks. The pole mass
$M^{}_t$ of the top quark can directly be measured. The relation
between the pole mass $M^{}_q$ and the $\overline{\rm MS}$ running
mass $m^{}_q (\mu)$ can be established by taking account of the
perturbative QCD corrections \cite{PDG}. Given the observed mass of
the Higgs boson $M^{}_H \simeq 125$ GeV, we have calculated the
running masses of six quarks at a number of typical energy scales
and listed their values in Table \ref{table2}  \cite{XZZ2012}, where
$\Lambda^{}_{\rm VS} \simeq 4\times 10^{12}$ GeV denotes the cutoff
scale of vacuum stability in the SM. Note that the values of the
pole masses $M^{}_q$ and running masses $m^{}_q(M^{}_q)$ themselves,
rather than the running masses $m^{}_q(\mu)$ at these mass scales,
are given in the last two rows of Table \ref{table2} for the sake of
comparison. But the pole masses of the three light quarks are not
listed, simply because the perturbative QCD calculation is not
reliable in that energy region \cite{XZZ2012}.
\begin{table}[t]
\renewcommand\arraystretch{1.3}
\tbl{Running quark masses at some typical energy scales in the SM,
including the Higgs mass $M^{}_H \simeq 125~{\rm GeV}$ and the
vacuum-stability cutoff scale $\Lambda^{}_{\rm VS} \simeq 4\times
10^{12}~{\rm GeV}$ \cite{XZZ2012}.
\vspace{0.2cm}}
{\begin{tabular}{c|cccccc} \hline\hline
$\mu$ & $m^{}_u ~ ({\rm MeV})$ &  $m^{}_d ~ ({\rm MeV})$ &
$m^{}_s ~ ({\rm MeV})$ & $m^{}_c ~ ({\rm GeV})$
& $m^{}_b ~ ({\rm GeV})$ & $m^{}_t ~ ({\rm GeV})$ \\
\hline
$m^{}_c(m^{}_c)$ & $2.79^{+0.83}_{-0.82}$ & $5.69^{+0.96}_{-0.95}$ &
$116^{+36}_{-24}$ & $1.29^{+0.05}_{-0.11}$ & $5.95^{+0.37}_{-0.15}$
& $385.7^{+8.1}_{-7.8}$\\
\hline
$2 ~{\rm GeV}$ & $2.4^{+0.7}_{-0.7}$ & $4.9 \pm 0.8$  &
$100^{+30}_{-20}$
& $1.11^{+0.07}_{-0.14}$ & $5.06^{+0.29}_{-0.11}$ & $322.2^{+5.0}_{-4.9}$ \\
\hline
$m^{}_b(m^{}_b)$ & $2.02^{+0.60}_{-0.60}$ & $4.12^{+0.69}_{-0.68}$ &
$84^{+26}_{-17}$ & $0.934^{+0.058}_{-0.120}$ &
$4.19^{+0.18}_{-0.16}$
& $261.8^{+3.0}_{-2.9}$ \\
\hline
$M^{}_W$ & $1.39^{+0.42}_{-0.41}$ & $2.85^{+0.49}_{-0.48}$ &
$58^{+18}_{-12}$
& $0.645^{+0.043}_{-0.085} $ & $2.90^{+0.16}_{-0.06}$ & $174.2 \pm 1.2$  \\
\hline
$M^{}_Z$ & $1.38^{+0.42}_{-0.41}$ & $2.82 \pm 0.48$ &
$57^{+18}_{-12}$
& $0.638^{+0.043}_{-0.084}$ & $2.86^{+0.16}_{-0.06}$ & $172.1 \pm 1.2 $ \\
\hline
$M^{}_H$ & $1.34^{+0.40}_{-0.40}$ & $2.74^{+0.47}_{-0.47}$ &
$56^{+17}_{-12}$ & $0.621^{+0.041}_{-0.082}$ & $2.79^{+0.15}_{-0.06}$ &
$167.0^{+1.2}_{-1.2}$ \\
\hline
$m^{}_t(m^{}_t)$ & $1.31^{+0.40}_{-0.39}$ & $2.68 \pm 0.46 $ &
$55^{+17}_{-11}$ & $0.608^{+0.041}_{-0.080}$ &
$2.73^{+0.15}_{-0.06}$
& $163.3 \pm 1.1$ \\
\hline
$1~{\rm TeV}$ & $1.17 \pm 0.35$ & $2.40^{+0.42}_{-0.41} $ &
$49^{+15}_{-10}$ & $0.543^{+0.037}_{-0.072}$ &
$2.41^{+0.14}_{-0.05}$
& $148.1 \pm 1.3$ \\
\hline
$\Lambda^{}_{\rm VS}$ & $0.61^{+0.19}_{-0.18}$ & $1.27 \pm 0.22$ &
$26^{+8}_{-5}$ & $0.281^{+0.02}_{-0.04}$ & $1.16^{+0.07}_{-0.02}$
& $82.6 \pm 1.4$\\
\hline \hline
$M^{}_q$ & --- & --- & --- & $1.84^{+0.07}_{-0.13}$
& $4.92^{+0.21}_{-0.08}$ & $172.9 \pm 1.1$ \\
\hline
$m^{}_q(M^{}_q)$ & --- & --- & --- & $1.14^{+0.06}_{-0.12}$
& $4.07^{+0.18}_{-0.06}$ & $162.5 \pm 1.1 $ \\
\hline\hline
\end{tabular}}
\label{table2}
\end{table}

The quark mass values shown in Table \ref{table2} indicate the
existence of a strong hierarchy either in the $(u, c, t)$ sector or
in the $(d, s, b)$ sector. We find that it is instructive to
consider the quark mass spectrum at the reference scale $\mu =
M^{}_Z$ by adopting the $\overline{\rm MS}$ scheme. The reason is
simply that an extension of the SM with new physics should be highly
necessary far above $M^{}_Z$, and the strong coupling constant
$\alpha^{}_{\rm s}$ becomes sizable far below $M^{}_Z$.
Quantitatively,
\begin{eqnarray}
Q^{}_q = +2/3: &~~& \frac{m^{}_u}{m^{}_c} \simeq
\frac{m^{}_c}{m^{}_t} \simeq \lambda^4 \; ,
\nonumber \\
Q^{}_q = -1/3: &~~& \frac{m^{}_d}{m^{}_s} \simeq
\frac{m^{}_s}{m^{}_b} \simeq \lambda^2 \; ,
\label{eq1}
\end{eqnarray}
hold to an acceptable degree of accuracy, where $\lambda \equiv
\sin\theta^{}_{\rm C} \approx 0.225$ with $\theta^{}_{\rm C}$ being
the famous Cabibbo angle of quark flavor mixing \cite{Cabibbo}.
The three charged leptons have a similar mass hierarchy.

\begin{figure}[t]
\centerline{\includegraphics[width=12cm]{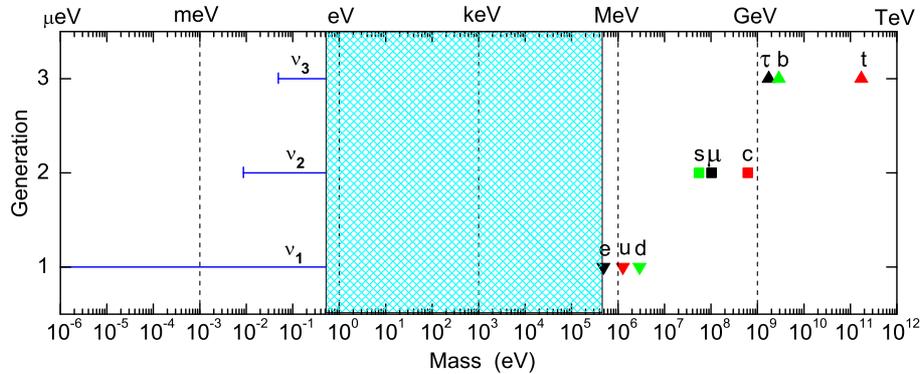}} \caption{A
schematic illustration of the flavor ``hierarchy" and ``desert"
problems in the SM fermion mass spectrum at the electroweak scale.}
\label{fig1}
\end{figure}

To be more intuitive, we present a schematic plot for the mass
spectrum of six quarks and six leptons at the electroweak scale in
Fig. \ref{fig1} \cite{Li}, where a normal neutrino mass ordering has
been assumed. One can see that the span between the neutrino masses
$m^{}_i$ and the top-quark mass $m^{}_t$ is at least twelve orders
of magnitude. Furthermore, the ``desert" between the heaviest
neutral fermion (e.g., $\nu^{}_3$) and the lightest charged fermion
(i.e., $e^-$) spans at least six orders of magnitude. Why do the
twelve fermions have such a strange mass pattern with the remarkable
hierarchy and desert? A convincing answer to this fundamental
question remains open.

\section{Flavor Mixing Pattern}

In a straightforward extension of the SM which allows its three
neutrinos to be massive, a nontrivial mismatch between the mass and
flavor eigenstates of leptons or quarks arises from the fact that
lepton or quark fields can interact with both scalar and gauge
fields, leading to the puzzling phenomena of flavor mixing and CP
violation. The $3\times 3$ lepton and quark flavor mixing matrices
appearing in the weak charged-current interactions are referred to,
respectively, as the Pontecorvo-Maki-Nakagawa-Sakata (PMNS) matrix
$U$ \cite{PMNS} and the Cabibbo-Kobayashi-Maskawa (CKM) matrix $V$
\cite{CKM}:
\begin{eqnarray}
-{\cal L}^{\ell}_{\rm cc} & = & \frac{g}{\sqrt{2}} \
\overline{\left(e ~~ \mu ~~ \tau \right)^{}_{\rm L}} \ \gamma^\mu \
U \left(\begin{matrix} \nu^{}_1 \cr \nu^{}_2 \cr \nu^{}_3
\cr \end{matrix}\right)^{}_{\rm L} W^-_\mu + {\rm h.c.} \; ,
\nonumber \\
-{\cal L}^{q}_{\rm cc} & = & \frac{g}{\sqrt{2}} \
\overline{\left(u ~~ c ~~ t \right)^{}_{\rm L}} \ \gamma^\mu \ V
\left(\begin{matrix} \hspace{0.08cm} d \hspace{0.08cm} \cr s \cr b
\cr \end{matrix}\right)^{}_{\rm L} W^+_\mu + {\rm h.c.} \; ,
\label{eq2}
\end{eqnarray}
in which all the fermion fields are the mass eigenstates. By
convention, $U$ and $V$ are defined to be associated with $W^-$ and
$W^+$, respectively. Note that $V$ is unitary as dictated by the SM
itself, but whether $U$ is unitary or not depends on the mechanism
responsible for the origin of neutrino masses.

In ${\cal L}^{\ell}_{\rm cc}$ and ${\cal L}^{q}_{\rm cc}$, the
charged leptons and quarks with the same electric charges all have
the normal mass hierarchies (namely, $m^{}_e \ll m^{}_\mu \ll
m^{}_\tau$, $m^{}_u \ll m^{}_c \ll m^{}_t$ and $m^{}_d \ll m^{}_s
\ll m^{}_b$, as shown in Fig. \ref{fig1} or Table \ref{table2}). Yet
it remains unclear whether the three neutrinos also have a normal
mass ordering ($m^{}_1 < m^{}_2 < m^{}_3$) or not. Now that $m^{}_1
< m^{}_2$ has been fixed from the solar neutrino oscillations, the
only likely ``abnormal" mass ordering is $m^{}_3 < m^{}_1 < m^{}_2$.
The neutrino mass ordering is one of the central concerns in flavor
physics, and it will be determined in the foreseeable future with
the help of either an accelerator-based neutrino oscillation
experiment or a reactor-based antineutrino oscillation experiment,
or both of them.

\begin{figure}[t]
\centerline{\includegraphics[width=12cm]{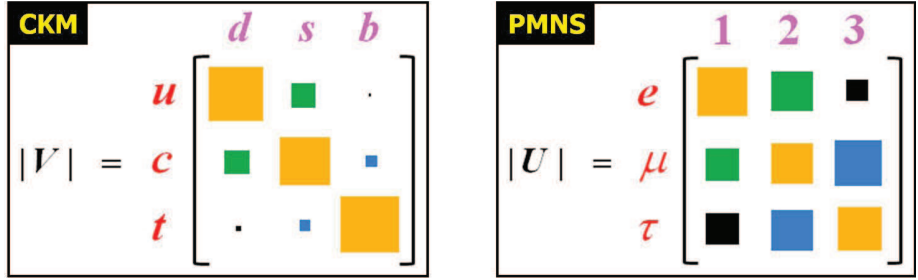}}
\vspace{0.15cm}
\caption{A schematic illustration of the different flavor mixing
structures of six quarks and six leptons at the electroweak scale.}
\label{fig2}
\end{figure}

Up to now the moduli of nine elements of the CKM matrix $V$ have
been determined from current experimental data to a good degree of
accuracy \cite{PDG}:
\begin{eqnarray}
|V| = \left(\begin{matrix} 0.97427 \pm 0.00015 & 0.22534 \pm 0.00065
& 0.00351^{+0.00015}_{-0.00014} \cr 0.22520 \pm 0.00065 & 0.97344
\pm 0.00016 & 0.0412^{+0.0011}_{-0.0005} \cr
0.00867^{+0.00029}_{-0.00031} & 0.0404^{+0.0011}_{-0.0005} &
0.999146^{+0.000021}_{-0.000046} \cr \end{matrix} \right) \; . ~~
\label{eq3}
\end{eqnarray}
We see that $V$ has a clear hierarchy: $|V^{}_{tb}| > |V^{}_{ud}| >
|V^{}_{cs}| \gg |V^{}_{us}| > |V^{}_{cd}| \gg |V^{}_{cb}| >
|V^{}_{ts}| \gg |V^{}_{td}| > |V^{}_{ub}|$, which must have
something to do with the strong hierarchy of quark masses. Fig.
\ref{fig2} illustrates the salient structural features of $V$, as
compared with the more or less ``anarchical" structure of the PMNS
matrix $U$. There exist at least two open questions \cite{Xing2013}:
a) Is there any intrinsic relationship between the flavor mixing
parameters of leptons and quarks in a certain grand unified theory?
b) If yes, does this kind of relationship hold between $U$ and $V$
or between $U$ and $V^\dagger$ (or between $U^\dagger$ and $V$)
which are both associated with $W^-$ (or $W^+$)?

The CKM matrix $V$ can be parametrized in terms of three flavor
mixing angles and a nontrivial CP-violating phase in nine different
ways \cite{FX98}. Among them, the most popular one is the so-called
``standard" parametrization as advocated by the Particle Data Group
\cite{PDG}:
\begin{eqnarray}
V = \left( \begin{matrix} c^{}_{12} c^{}_{13} & s^{}_{12} c^{}_{13}
& s^{}_{13} e^{-{\rm i} \delta^{}_q} \cr -s^{}_{12} c^{}_{23} -
c^{}_{12} s^{}_{13} s^{}_{23} e^{{\rm i} \delta^{}_q} & c^{}_{12}
c^{}_{23} - s^{}_{12} s^{}_{13} s^{}_{23} e^{{\rm i} \delta^{}_q} &
c^{}_{13} s^{}_{23} \cr s^{}_{12} s^{}_{23} - c^{}_{12} s^{}_{13}
c^{}_{23} e^{{\rm i} \delta^{}_q} & -c^{}_{12} s^{}_{23} - s^{}_{12}
s^{}_{13} c^{}_{23} e^{{\rm i} \delta^{}_q} & c^{}_{13} c^{}_{23}
\cr
\end{matrix} \right) \; ,
\label{eq4}
\end{eqnarray}
in which $c^{}_{ij} \equiv \cos\vartheta^{}_{ij}$ and $s^{}_{ij}
\equiv \sin\vartheta^{}_{ij}$ (for $ij = 12, 13, 23$) are
defined. The present experimental data lead us to
\begin{eqnarray}
\vartheta^{}_{12} = 13.023^\circ \pm 0.038^\circ \; , ~~
\vartheta^{}_{13} = 0.201^{+0.009^\circ}_{-0.008^\circ} \; , ~~
\vartheta^{}_{23} = 2.361^{+0.063^\circ}_{-0.028^\circ} \; , ~~
\label{eq5}
\end{eqnarray}
and $\delta^{}_q = 69.21^{+2.55^\circ}_{-4.59^\circ}$. In
comparison, the similar parameters of the PMNS lepton flavor mixing
matrix $U$ lie in the following $3\sigma$ ranges as obtained from a
global analysis of current neutrino oscillation data \cite{Fogli}:
\begin{eqnarray}
\theta^{}_{12} = 30.6^\circ \to 36.8^\circ \; , ~~
\theta^{}_{13} = 7.6^\circ \to 9.9^\circ \; , ~~
\theta^{}_{23} = 37.7^\circ \to 52.3^\circ \; , ~~
\label{eq6}
\end{eqnarray}
and $\delta^{}_\ell = 0^\circ \to 360^\circ$ provided the neutrino
mass ordering is normal (i.e., $m^{}_1 < m^{}_2 < m^{}_3$); or
\begin{eqnarray}
\theta^{}_{12} = 30.6^\circ \to 36.8^\circ \; , ~~
\theta^{}_{13} = 7.7^\circ \to 9.9^\circ \; , ~~
\theta^{}_{23} = 38.1^\circ \to 53.2^\circ \; , ~~
\label{eq7}
\end{eqnarray}
and $\delta^{}_\ell = 0^\circ \to 360^\circ$ provided the neutrino
mass ordering is inverted (i.e., $m^{}_3 < m^{}_1 < m^{}_2$). In
either case $U$ exhibits an anarchical pattern as shown in Fig.
\ref{fig2} (left panel). In the literature the possibilities of
$\theta^{}_{12} + \vartheta^{}_{12} = 45^\circ$ and $\theta^{}_{23}
\pm \vartheta^{}_{23} = 45^\circ$ have been discussed, although such
relations depend on both the chosen parametrization and the chosen
energy scale \cite{Xing05}.

It is worth mentioning the off-diagonal asymmetries of the CKM
matrix $V$ in modulus \cite{Xing95}, which provide another measure
of the structure of $V$ about its
$V^{}_{ud}$-$V^{}_{cs}$-$V^{}_{tb}$ and
$V^{}_{ub}$-$V^{}_{cs}$-$V^{}_{td}$ axes, respectively:
\begin{eqnarray}
\Delta^{q}_{\rm L} & \equiv & |V^{}_{us}|^2 - |V^{}_{cd}|^2 =
|V^{}_{cb}|^2 - |V^{}_{ts}|^2 = |V^{}_{td}|^2 - |V^{}_{ub}|^2 \simeq
A^2 \lambda^6 \left(1 - 2\rho\right) \; ,
\nonumber \\
\Delta^{q}_{\rm R} & \equiv & |V^{}_{us}|^2 - |V^{}_{cb}|^2 =
|V^{}_{cd}|^2 - |V^{}_{ts}|^2 = |V^{}_{tb}|^2 - |V^{}_{ud}|^2 \simeq
\lambda^2 \; ,
\label{eq8}
\end{eqnarray}
where $A \simeq 0.811$, $\lambda \simeq 0.225$ and $\rho \simeq
0.134$ denote the so-called Wolfenstein parameters \cite{Wol}. It
becomes obvious that $\Delta^q_{\rm L} \simeq 6.3 \times 10^{-5}$
and $\Delta^q_{\rm R} \simeq 5.1 \times 10^{-2}$ hold, implying that
the CKM matrix $V$ is symmetric about its
$V^{}_{ud}$-$V^{}_{cs}$-$V^{}_{tb}$ axis to a high degree of
accuracy. In comparison, the PMNS matrix $U$ is not that symmetric
about its either axis, but it may possess an approximate or partial
$\mu$-$\tau$ permutation symmetry \cite{XZ2014}; i.e., $|U^{}_{\mu
i}| \simeq |U^{}_{\tau i}|$ (for $i=1,2,3$). Such an interesting
lepton flavor mixing structure at low energies might originate, via
the renormalization-group running effects, from a superhigh-energy
PMNS matrix with the exact $\mu$-$\tau$ symmetry \cite{LX2014}.

\section{The Unitarity Triangles}

\begin{figure}[t]
\vspace{-1cm}
\centerline{\hspace{0.5cm}\includegraphics[width=16cm]{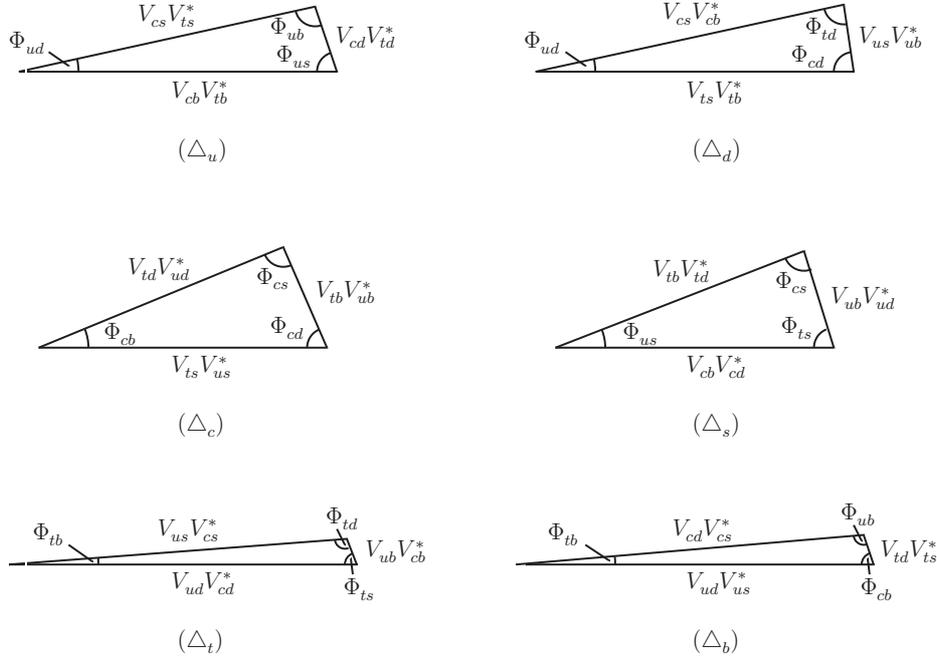}}
\vspace{-10.8cm}
\caption{A schematic illustration of the six CKM unitarity
triangles in the complex plane, where each triangle is named
by the flavor index that does not manifest itself in the sides.}
\label{fig3}
\end{figure}

Thanks to the six orthogonality relations of the unitary CKM matrix
$V$, one may define six unitarity triangles in the complex plane:
\begin{eqnarray}
\triangle^{}_\alpha : &~~~& V^{}_{\beta d} V^*_{\gamma d} +
V^{}_{\beta s} V^*_{\gamma s} + V^{}_{\beta b} V^*_{\gamma b} = 0 \; ,
\nonumber \\
\triangle^{}_i : &~~~& V^{}_{u j} V^*_{u k} + V^{}_{c j} V^*_{c k} +
V^{}_{t j} V^*_{t k} = 0 \; ,
\label{eq9}
\end{eqnarray}
where $\alpha$, $\beta$ and $\gamma$ co-cyclically run over the
up-type quarks $u$, $c$ and $t$, while $i$, $j$ and $k$
co-cyclically run over the down-type quarks $d$, $s$ and $b$. The
inner angles of triangles $\triangle^{}_\alpha$ and $\triangle^{}_i$
are universally defined as
\begin{eqnarray}
\Phi^{}_{\alpha i} \equiv \arg\left(- \frac{V^{}_{\beta j}
V^*_{\gamma j}} {V^{}_{\beta k} V^*_{\gamma k}} \right) =
\arg\left(- \frac{V^{}_{\beta j} V^*_{\beta k}} {V^{}_{\gamma j}
V^*_{\gamma k}} \right) \; , \label{eq10}
\end{eqnarray}
where the Greek and Latin subscripts keep their separate co-cyclical
running. So $\triangle^{}_\alpha$ and $\triangle^{}_i$ share a
common inner angle $\Phi^{}_{\alpha i}$, as shown in Fig. \ref{fig3}.

Let us proceed to define the Jarlskog invariant of CP violation
${\cal J}^{}_q$ for the CKM matrix $V$ through the equation \cite{J}
\begin{eqnarray}
{\rm Im}\left( V^{}_{\alpha i} V^{}_{\beta j} V^*_{\alpha j}
V^*_{\beta i} \right) = {\cal J}^{}_q \sum_{\gamma}
\epsilon^{}_{\alpha\beta\gamma} \sum_{k} \epsilon^{}_{ijk} \; ,
\label{eq11}
\end{eqnarray}
where the relevant Greek and Latin subscripts run over $(u, c, t)$
and $(d, s, b)$, respectively. Given the standard (or Wolfenstein)
parametrization of $V$,
\begin{eqnarray}
{\cal J}^{}_q = c^{}_{12} s^{}_{12} c^2_{13} s^{}_{13} c^{}_{23}
s^{}_{23} \sin\delta^{}_q \simeq A^2 \lambda^6 \eta \; ,
\label{eq12}
\end{eqnarray}
where $\eta \simeq 0.354$ is the fourth Wolfenstein parameter
\cite{Wol}. We are therefore left with ${\cal J}^{}_q \simeq 3.0
\times 10^{-5}$, comparable with $\Delta^q_{\rm L}$ in magnitude.
The six CKM unitarity triangles have the same area, equal to ${\cal
J}^{}_q/2$. If $\Delta^q_{\rm L}$ or $\Delta^q_{\rm R}$ were
vanishing, there would be the congruence between two unitarity
triangles \cite{FX00}:
\begin{eqnarray}
\Delta^q_{\rm L} =0 & ~ \Longrightarrow ~ & \triangle^{}_u \cong
\triangle^{}_d \; , ~~ \triangle^{}_c \cong \triangle^{}_s \; , ~~
\triangle^{}_t \cong \triangle^{}_b \; ,
\nonumber \\
\Delta^q_{\rm R} =0 & \Longrightarrow & \triangle^{}_u \cong
\triangle^{}_b \; , ~~ \triangle^{}_c \cong \triangle^{}_s \; , ~~
\triangle^{}_t \cong \triangle^{}_d \; .
\label{eq13}
\end{eqnarray}
Fig. \ref{fig3} clearly shows that $\Delta^q_{\rm L} \simeq 0$
is actually a rather good approximation.

Note that triangle $\triangle^{}_s$ has been well studied in
$B$-meson physics, and its three inner angles are usually denoted as
$\alpha = \Phi^{}_{cs}$, $\beta = \Phi^{}_{us}$ and $\gamma =
\Phi^{}_{ts}$. A very striking result is $\alpha =
89.0^{+4.4^\circ}_{-4.2^\circ}$ as reported by the Particle Data
Group \cite{PDG}, implying that triangles $\triangle^{}_s$ and
$\triangle^{}_c$ are almost the {\it right} triangles. In fact,
$\alpha =90^\circ$ leads us to the parameter correlation ${\cal
J}^{}_{\rm q} = |V^{}_{ud}| \cdot |V^{}_{ub}| \cdot |V^{}_{td}|
\cdot |V^{}_{tb}|$, or equivalently $\cos\delta^{}_q =
\sin\vartheta^{}_{13}/(\tan\vartheta^{}_{12} \tan\vartheta^{}_{23})$
or $\eta \simeq \sqrt{\rho (1 - \rho)}$. If the CKM matrix $V$ is
parametrized as \cite{FX97}
\begin{eqnarray}
V = \left( \begin{matrix} s^{}_{\rm u} s^{}_{\rm d} c^{}_{\rm h} +
c^{}_{\rm u} c^{}_{\rm d} e^{-{\rm i}\phi} & ~ s^{}_{\rm u}
c^{}_{\rm d} c^{}_{\rm h} - c^{}_{\rm u} s^{}_{\rm d} e^{-{\rm
i}\phi} ~ & s^{}_{\rm u} s^{}_{\rm h} \cr c^{}_{\rm u} s^{}_{\rm d}
c^{}_{\rm h} - s^{}_{\rm u} c^{}_{\rm d} e^{-{\rm i}\phi} &
c^{}_{\rm u} c^{}_{\rm d} c^{}_{\rm h} + s^{}_{\rm u} s^{}_{\rm d}
e^{-{\rm i}\phi} & c^{}_{\rm u} s^{}_{\rm h} \cr -s^{}_{\rm d}
s^{}_{\rm h} & -c^{}_{\rm d} s^{}_{\rm h} & c^{}_{\rm h} \cr
\end{matrix} \right) \; ,
\label{eq14}
\end{eqnarray}
where $c^{}_x \equiv \cos\vartheta^{}_x$ and $s^{}_x \equiv
\sin\vartheta^{}_x$ with $x =$ u (up), d (down) or h (heavy),
then one can easily obtain the following
relationship \cite{Xing2009}:
\begin{equation}
\frac{\sin \alpha}{\sin\phi} \simeq 1 - \tan\vartheta^{}_{\rm u}
\tan\vartheta^{}_{\rm d} \cos\vartheta^{}_{\rm h} \cos\phi -
\frac{1}{2} \tan^2\vartheta_{\rm u} \tan^2\vartheta_{\rm d}
\cos^2\vartheta^{}_{\rm h} \; ,
\label{eq15}
\end{equation}
where the higher-order terms of $\tan\vartheta^{}_{\rm u}$ and
$\tan\vartheta^{}_{\rm d}$ have been omitted. So $\alpha \simeq
\phi$ holds to an excellent degree of accuracy. Taking account of
\begin{eqnarray}
\vartheta^{}_{\rm u} & = & \arctan\left(\left|
\frac{V^{}_{ub}}{V^{}_{cb}}\right|\right) \simeq 4.87^\circ \; ,
\nonumber \\
\vartheta^{}_{\rm d} & = & \arctan\left(\left|
\frac{V^{}_{td}}{V^{}_{ts}}\right|\right) \simeq 12.11^\circ \; ,
\nonumber \\
\vartheta^{}_{\rm h} & = & \arcsin\left(\sqrt{|V^{}_{ub}|^2 +
|V^{}_{cb}|^2}\right) \simeq 2.37^\circ \; ,
\label{eq16}
\end{eqnarray}
we obtain $\alpha \simeq 88.95^\circ$ from $\phi = 90^\circ$
according to Eq. (\ref{eq15}). The numerical result $\phi - \alpha
\simeq 1.05^\circ$ is extremely interesting in the sense that our
current experimental data strongly hint at $\phi = 90^\circ$ either
at the electroweak scale or at a superhigh-energy scale. Such a
conclusion holds because it has been proved that both $\phi$ and
$\alpha$ are completely insensitive to the renormalization-group
running effects \cite{Xing2009,LX2010}. Note also that $\phi$
essentially measures the phase difference between the up-type quark
mass matrix and the down-type quark mass matrix, and thus $\phi =
90^\circ$ might have very profound meaning with respect to the
origin of CP violation. Some authors have also discussed the possibilities
of geometrical, maximal or minimal CP violation \cite{Gerard} in
the quark sector.

Finally let us make a brief comment on the so-called CKM phase
matrix, whose elements are just the nine inner angles of the CKM
unitarity triangles as defined in Eq. (\ref{eq10})
\cite{LX2010,Harrison}:
\begin{eqnarray}
\Phi = \left ( \begin{matrix} \Phi^{}_{ud} & ~\Phi^{}_{us}~ &
\Phi^{}_{ub} \cr \Phi^{}_{cd} & \Phi^{}_{cs} & \Phi^{}_{cb} \cr
\Phi^{}_{td} & \Phi^{}_{ts} & \Phi^{}_{tb} \end{matrix} \right)
\simeq \left ( \begin{matrix} 1.05^\circ & 21.38^\circ &
157.57^\circ \cr 68.65^\circ & ~~ 88.95^\circ ~~ & 22.4^\circ \cr
110.3^\circ & 69.67^\circ & 0.034^\circ \end{matrix} \right) \; .
\label{eq17}
\end{eqnarray}
Each row or column of $\Phi$ corresponds to an explicit unitarity
triangle as illustrated in Fig. \ref{fig3}, and thus its three matrix
elements must satisfy the following six sum rules:
\begin{eqnarray}
\sum_\alpha \Phi^{}_{\alpha i} = \sum_i \Phi^{}_{\alpha i} =
180^\circ \; .
\label{eq18}
\end{eqnarray}
So one may similarly define two off-diagonal asymmetries of $\Phi$
about its two axes \cite{LX2010}. If one of the two asymmetries were
vanishing, we would be left with a result which is analogous to the
one in Eq. (\ref{eq13}). Of course, one may easily extend the same
language to describe the PMNS phase matrix and discuss its evolution
with the energy scales \cite{Luo}.

\section{Two Important Limits}

Now let us speculate whether the observed pattern of the CKM matrix
$V$ can be partly understood in some reasonable limits of quark
masses. This idea is more or less motivated by two useful working
symmetries in understanding the strong interactions of quarks and
hadrons by means of QCD or an effective field theory based on QCD
\cite{W}: the chiral quark symmetry (i.e., $m^{}_u, m^{}_d, m^{}_s
\to 0$) and the heavy quark symmetry (i.e., $m^{}_c, m^{}_b, m^{}_t
\to \infty$). The reason for the usefulness of these two symmetries
is simply that the masses of the light quarks are far below the
typical QCD scale $\Lambda^{}_{\rm QCD} \sim 0.2$ GeV, whereas the
masses of the heavy quarks are far above it. Because the elements of
$V$ are dimensionless and their magnitudes lie in the range of 0 to
1, they can only depend on the mass ratios of the lighter quarks to
the heavier quarks. The mass limits corresponding to the chiral and
heavy quark symmetries are therefore equivalent to setting the
relevant mass ratios to zero, and they are possible to help reveal a
part of the salient features of $V$. In this spirit, some
preliminary attempts have been made to look at the quark flavor
mixing pattern in the $m^{}_u, m^{}_d \to 0$ or $m^{}_t, m^{}_b
\to\infty$ limits \cite{F87}.

We shall show that it is possible to gain an insight into the
observed pattern of quark flavor mixing in the chiral and heavy
quark mass limits. This model-independent access to the underlying
quark flavor structure can explain why $|V^{}_{us}| \simeq
|V^{}_{cd}|$ and $|V^{}_{cb}| \simeq |V^{}_{ts}|$ hold to a good
degree of accuracy, why $|V^{}_{cd}/V^{}_{td}| \simeq
|V^{}_{cs}/V^{}_{ts}| \simeq |V^{}_{tb}/V^{}_{cb}|$ is a reasonable
approximation, and why $|V^{}_{ub}/V^{}_{cb}|$ should be smaller
than $|V^{}_{td}/V^{}_{ts}|$. Furthermore, the empirical relations
$|V^{}_{ub}/V^{}_{cb}| \sim \sqrt{m^{}_u/m^{}_c}$ and
$|V^{}_{td}/V^{}_{ts}| \simeq \sqrt{m^{}_d/m^{}_s}$ can be
reasonably conjectured in the heavy quark mass limits
\cite{Xing2012}.

Let us begin with the CKM matrix $V = O^\dagger_{\rm u} O^{}_{\rm
d}$ with $O^{}_{\rm u}$ and $O^{}_{\rm d}$ being the unitary
transformations responsible for the diagonalizations of the up- and
down-type quark mass matrices in the flavor basis. Namely,
\begin{eqnarray}
O^\dagger_{\rm u} H^{}_{\rm u} O^{}_{\rm u} & = &
O^\dagger_{\rm u} M^{}_{\rm u} M^\dagger_{\rm u}
O^{}_{\rm u} = {\rm Diag} \left\{m^2_u , m^2_c , m^2_t \right\} \; ,
\nonumber \\
O^\dagger_{\rm d} H^{}_{\rm d} O^{}_{\rm d} & = &
O^\dagger_{\rm d} M^{}_{\rm d} M^\dagger_{\rm d}
O^{}_{\rm d} = {\rm Diag} \left\{m^2_d , m^2_s , m^2_b \right\} \; ,
\label{eq19}
\end{eqnarray}
where $H^{}_{\rm u}$ and $H^{}_{\rm d}$ are defined to be Hermitian.
To be more explicit, the nine matrix elements of $V$ read
\begin{eqnarray}
V^{}_{\alpha i} & = & \sum^3_{k=1}
(O^{}_{\rm u})^*_{k \alpha} (O^{}_{\rm d})^{}_{k i} \; , ~~
\label{eq20}
\end{eqnarray}
where $\alpha$ and $i$ run over $(u, c, t)$ and $(d, s, b)$,
respectively. In general, the mass limit $m^{}_u \to 0$ (or $m^{}_d
\to 0$) does not correspond to a unique form of $H^{}_{\rm u}$ (or
$H^{}_{\rm d}$). The reason is simply that the form of a fermion
mass matrix is always basis-dependent. Without loss of any
generality, one may choose a particular flavor basis such that
$H^{}_{\rm u}$ and $H^{}_{\rm d}$ can be written as
\begin{eqnarray}
\lim_{m^{}_u \to 0} H^{}_{\rm u} & = & \left( \begin{matrix} 0 & 0 &
0 \cr 0 & ~ \times ~ & \times \cr 0 & \times & \times \cr
\end{matrix} \right) \; ,
\nonumber \\
\lim_{m^{}_d \to 0} H^{}_{\rm d} & = & \left( \begin{matrix} 0 & 0 &
0 \cr 0 & ~ \times ~ & \times \cr 0 & \times & \times \cr
\end{matrix} \right) \; ,
\label{eq21}
\end{eqnarray}
in which ``$\times$" denotes an arbitrary  nonzero element. Note
that Eq. (\ref{eq21}) is the result of a basis choice instead of an
assumption \cite{Xing2012}. When the mass of a given quark goes to
infinity, we argue that it becomes decoupled from the masses of
other quarks. In this case one may also choose a specific flavor
basis where $H^{}_{\rm u}$ and $H^{}_{\rm d}$ can be written as
\begin{eqnarray}
\lim_{m^{}_t \to \infty} H^{}_{\rm u} & = & \left( \begin{matrix}
\times & ~ \times ~ & 0 \cr \times & \times & 0 \cr 0 & 0 & \infty
\cr \end{matrix} \right) \; ,
\nonumber \\
\lim_{m^{}_b \to \infty} H^{}_{\rm d} & = & \left( \begin{matrix}
\times & ~ \times ~ & 0 \cr \times & \times & 0 \cr 0 & 0 & \infty
\cr \end{matrix} \right) \; .
\label{eq22}
\end{eqnarray}
In other words, the $3\times 3$ Hermitian matrices $H^{}_{\rm u}$
and $H^{}_{\rm d}$ can be simplified to the effective $2\times 2$
Hermitian matrices in either the chiral quark mass limit or the
heavy quark mass limit. In view of the fact that $m^{}_u \ll m^{}_c
\ll m^{}_t$ and $m^{}_d \ll m^{}_s \ll m^{}_b$ hold at an arbitrary
energy scale, as shown in Table \ref{table2}, we believe that Eqs.
(\ref{eq21}) and (\ref{eq22}) are phenomenologically reasonable and
can help explain some of the observed properties of quark flavor
mixing in a model-independent way. Let us go into details.

(1) {\it Why $|V^{}_{us}| \simeq |V^{}_{cd}|$ and $|V^{}_{cb}|
\simeq |V^{}_{ts}|$ hold?} --- A glance at Eq. (\ref{eq3}) tells us
that $|V^{}_{us}| \simeq |V^{}_{cd}|$ is an excellent approximation.
It can be well understood in the heavy quark mass limits, where
Hermitian $H^{}_{\rm u}$ and $H^{}_{\rm d}$ may take the form of Eq.
(\ref{eq22}). In this case the unitary matrices $O^{}_{\rm u}$ and
$O^{}_{\rm d}$ used to diagonalize $H^{}_{\rm u}$ and $H^{}_{\rm d}$
can be expressed as
\begin{eqnarray}
\lim_{m^{}_t \to \infty} O^{}_{\rm u} & = & P^{}_{12} \left(
\begin{matrix} c^{}_{12} & s^{}_{12} & 0 \cr -s^{}_{12} & c^{}_{12}
& 0 \cr 0 & 0 & 1 \cr \end{matrix} \right) \; ,
\nonumber \\
\lim_{m^{}_b \to \infty} O^{}_{\rm d} & = & P^\prime_{12} \left(
\begin{matrix} c^{\prime}_{12} & s^{\prime}_{12} & 0 \cr
-s^{\prime}_{12} & c^{\prime}_{12} & 0 \cr 0 & 0 & 1 \cr
\end{matrix} \right) \; ,
\label{eq23}
\end{eqnarray}
where $c^{(\prime)}_{12} \equiv \cos\vartheta^{(\prime)}_{12}$,
$s^{(\prime)}_{12} \equiv \sin\vartheta^{(\prime)}_{12}$, and
$P^{(\prime)}_{12} = {\rm Diag}\left\{e^{i\phi^{(\prime)}_{12}}, 1,
1 \right\}$. Therefore, we immediately arrive at
\begin{eqnarray}
\left|V^{}_{us}\right| = \left|c^{}_{12} s^\prime_{12} - s^{}_{12}
c^\prime_{12} e^{i\Delta^{}_{12}}\right| = \left|V^{}_{cd}\right|
\label{eq24}
\end{eqnarray}
in the $m^{}_t \to \infty$ and $m^{}_b \to \infty$ limits, where
$\Delta^{}_{12} \equiv \phi^\prime_{12} - \phi^{}_{12}$ denotes the
nontrivial phase difference between the up- and down-quark sectors.
Since $m^{}_u/m^{}_c \sim m^{}_c/m^{}_t \sim \lambda^4$ and
$m^{}_d/m^{}_s \sim m^{}_s/m^{}_b \sim \lambda^2$ hold, the mass
limits taken above are surely a good approximation. So the
approximate equality $|V^{}_{us}| \simeq |V^{}_{cd}|$ is naturally
attributed to the fact that both $m^{}_t \gg m^{}_u, m^{}_c$ and
$m^{}_b \gg m^{}_d, m^{}_s$ hold
\footnote{Quantitatively, $|V^{}_{us}| \simeq |V^{}_{cd}| \simeq
\lambda$ holds. Hence $s^{}_{12} \simeq \sqrt{m^{}_u/m^{}_c} \simeq
\lambda^2$ and $s^\prime_{12} \simeq \sqrt{m^{}_d/m^{}_s} \simeq
\lambda$ are often conjectured and can easily be derived from some
ans$\rm\ddot{a}$tze of quark mass matrices \cite{F77}.}.

One may similarly consider the chiral quark mass limits $m^{}_u \to
0$ and $m^{}_d \to 0$ so as to understand why $|V^{}_{ts}| \simeq
|V^{}_{cb}|$ holds. Eq. (\ref{eq21}) leads us to
\begin{eqnarray}
\lim_{m^{}_u \to 0} O^{}_{\rm u} & = & P^{}_{23} \left(
\begin{matrix} 1 & 0 & 0 \cr 0 & c^{}_{23} & s^{}_{23} \cr 0 &
-s^{}_{23} & c^{}_{23} \cr \end{matrix} \right) \; ,
\nonumber \\
\lim_{m^{}_d \to 0} O^{}_{\rm d} & = & P^\prime_{23} \left(
\begin{matrix} 1 & 0 & 0 \cr 0 & c^{\prime}_{23} & s^{\prime}_{23}
\cr 0 & -s^{\prime}_{23} & c^{\prime}_{23} \cr
\end{matrix} \right) \; ,
\label{eq25}
\end{eqnarray}
in which $c^{(\prime)}_{23} \equiv \cos\vartheta^{(\prime)}_{23}$,
$s^{(\prime)}_{23} \equiv \sin\vartheta^{(\prime)}_{23}$, and
$P^{(\prime)}_{23} = {\rm Diag}\left\{1, 1,
e^{i\phi^{(\prime)}_{23}}\right\}$. We are therefore left with
\begin{eqnarray}
\left|V^{}_{cb}\right| = \left|c^{}_{23} s^\prime_{23} - s^{}_{23}
c^\prime_{23} e^{i\Delta^{}_{23}} \right| = \left|V^{}_{ts}\right|
\label{eq26}
\end{eqnarray}
in the $m^{}_u \to 0$ and $m^{}_d \to 0$ limits, where
$\Delta^{}_{23} \equiv \phi^\prime_{23} - \phi^{}_{23}$ stands for
the nontrivial phase difference between the up- and down-quark
sectors. This model-independent result is also in good agreement
with the experimental data $|V^{}_{cb}| \simeq |V^{}_{ts}|$ as given
in Eq. (\ref{eq3}). Namely, the approximate equality $|V^{}_{cb}|
\simeq |V^{}_{ts}|$ is a natural consequence of $m^{}_u \ll m^{}_c,
m^{}_t$ and $m^{}_d \ll m^{}_s, m^{}_b$ in no need of any specific
assumptions
\footnote{It is possible to obtain the quantitative relationship
$|V^{}_{cb}| \simeq |V^{}_{ts}| \simeq \lambda^2$ through $s^{}_{23}
\simeq m^{}_c/m^{}_t \simeq \lambda^4$ and $s^\prime_{23} \simeq
m^{}_s/m^{}_b \simeq \lambda^2$ from a number of ans$\rm\ddot{a}$tze
of quark mass matrices \cite{FX00}.}.

(2) {\it Why $|V^{}_{cd}/V^{}_{td}| \simeq |V^{}_{cs}/V^{}_{ts}|
 \simeq |V^{}_{tb}/V^{}_{cb}|$ holds?} ---
Given the magnitudes of the CKM matrix elements in Eq. (\ref{eq3}),
it is easy to get $|V^{}_{cd}/V^{}_{td}| \simeq 26.0$,
$|V^{}_{cs}/V^{}_{ts}| \simeq 24.1$ and $|V^{}_{tb}/V^{}_{cb}|
\simeq 24.3$. Thus $|V^{}_{cd}/V^{}_{td}| \simeq
|V^{}_{cs}/V^{}_{ts}| \simeq |V^{}_{tb}/V^{}_{cb}|$ holds as a reasonably
good approximation. We find that such an approximate relation becomes
exact in the mass limits $m^{}_u \to 0$ and $m^{}_b \to \infty$. To
be much more explicit, we arrive at
\begin{eqnarray}
V = \lim_{m^{}_u \to 0} O^{\dagger}_{\rm u} \lim_{m^{}_b \to \infty}
O^{}_{\rm d} = P^\prime_{12} \left(\begin{matrix} c^\prime_{12} &
s^\prime_{12} & 0 \cr -c^{}_{23} s^\prime_{12} & c^{}_{23}
c^\prime_{12} & -s^{}_{23} \cr -s^{}_{23} s^\prime_{12} & s^{}_{23}
c^\prime_{12} & c^{}_{23} \cr
\end{matrix} \right) P^\dagger_{23} \; ,
\label{eq27}
\end{eqnarray}
where Eqs. (\ref{eq23}) and (\ref{eq25}) have been used. Therefore,
\begin{eqnarray}
\left|\frac{V^{}_{cd}}{V^{}_{td}}\right| =
\left|\frac{V^{}_{cs}}{V^{}_{ts}}\right| =
\left|\frac{V^{}_{tb}}{V^{}_{cb}}\right| =
\left|\cot\vartheta^{}_{23}\right| \;
\label{eq28}
\end{eqnarray}
holds in the chosen quark mass limits, which assure the smallest CKM
matrix element $V^{}_{ub}$ to vanish. This simple result is
essentially consistent with the experimental data if
$\vartheta^{}_{23} \simeq 2.35^\circ$ is taken
\footnote{This numerical estimate implies $\tan\vartheta^{}_{23}
\simeq \lambda^2 \simeq \sqrt{m^{}_c/m^{}_t}$, which can easily be
derived from the Fritzsch ansatz of quark mass matrices \cite{F78}.}.
Note that the quark mass limits $m^{}_t \to \infty$ and $m^{}_d \to
0$ are less favored because they predict both $|V^{}_{td}| =0$ and
$|V^{}_{us}/V^{}_{ub}| = |V^{}_{cs}/V^{}_{cb}| =
|V^{}_{tb}/V^{}_{ts}|$, which are in conflict with current
experimental data. In particular, the limit $|V^{}_{ub}| =0$ is
apparently closer to reality than the limit $|V^{}_{td}| =0$. But
why $V^{}_{ub}$ is smaller in magnitude than all the other CKM
matrix elements remains a puzzle, since it is difficult for us to
judge that the quark mass limits $m^{}_u \to \infty$ and $m^{}_b \to
0$ should make more sense than the quark mass limits $m^{}_t \to
\infty$ and $m^{}_d \to 0$ from a phenomenological point of view.
The experimental data in Eq. (\ref{eq3}) indicate $|V^{}_{td}|
\gtrsim 2|V^{}_{ub}|$ and $|V^{}_{ts}| \simeq |V^{}_{cb}|$. So a
comparison between the ratios $|V^{}_{ub}/V^{}_{cb}|$ and
$|V^{}_{td}/V^{}_{ts}|$ might be able to tell us an acceptable
reason for $|V^{}_{td}| > |V^{}_{ub}|$.

(3) {\it Why $|V^{}_{ub}/V^{}_{cb}|$ is smaller than
$|V^{}_{td}/V^{}_{ts}|$?} --- Given Eqs. (\ref{eq21}), (\ref{eq22})
and (\ref{eq23}), we can calculate the ratios
$|V^{}_{ub}/V^{}_{cb}|$ and $|V^{}_{td}/V^{}_{ts}|$ in the
respective heavy quark mass limits \cite{Xing2012}:
\begin{eqnarray}
\lim_{m^{}_b \to \infty} \left|\frac{V^{}_{ub}}{V^{}_{cb}}\right| &
= & \left|\frac{(O^{}_{\rm u})^{}_{3 u}}{(O^{}_{\rm u})^{}_{3
c}}\right| \; ,
\nonumber \\
\lim_{m^{}_t \to \infty} \left|\frac{V^{}_{td}}{V^{}_{ts}}\right| &
= & \left|\frac{(O^{}_{\rm d})^{}_{3 d}}{(O^{}_{\rm d})^{}_{3
s}}\right| \; . ~~
\label{eq29}
\end{eqnarray}
This result is quite nontrivial in the sense that
$|V^{}_{ub}/V^{}_{cb}|$ turns out to be independent of the mass
ratios of three down-type quarks in the $m^{}_b \to \infty$ limit,
and $|V^{}_{td}/V^{}_{ts}|$ has nothing to do with the mass ratios
of three up-type quarks in the $m^{}_t \to \infty$ limit. In
particular, the flavor indices showing up on the right-hand side of
Eq. (\ref{eq29}) is rather suggestive: $|V^{}_{ub}/V^{}_{cb}|$ is
relevant to $u$ and $c$ quarks, and $|V^{}_{td}/V^{}_{ts}|$ depends
on $d$ and $s$ quarks. We are therefore encouraged to conjecture
that $|V^{}_{ub}/V^{}_{cb}|$ (or $|V^{}_{td}/V^{}_{ts}|$) should be
a simple function of the mass ratio $m^{}_u/m^{}_c$ (or
$m^{}_d/m^{}_s$) in the $m^{}_t \to \infty$ (or $m^{}_b \to \infty$)
limit. If the values of $m^{}_u$, $m^{}_d$, $m^{}_s$ and $m^{}_c$ in
Table \ref{table2} are taken into account, the simplest
phenomenological conjectures should be
\begin{eqnarray}
\lim_{m^{}_b \to \infty} \left|\frac{V^{}_{ub}}{V^{}_{cb}}\right| &
\simeq & c^{}_1 \sqrt{\frac{m^{}_u}{m^{}_c}} \; ,
\nonumber \\
\lim_{m^{}_t \to \infty} \left|\frac{V^{}_{td}}{V^{}_{ts}}\right| &
\simeq & c^{}_2 \sqrt{\frac{m^{}_d}{m^{}_s}} \; ,
\label{eq30}
\end{eqnarray}
where $c^{}_1$ and $c^{}_2$ are the coefficients of ${\cal O}(1)$.
In view of $\sqrt{m^{}_u/m^{}_c} \simeq \lambda^2$ and
$\sqrt{m^{}_d/m^{}_s} \simeq \lambda$, we expect that
$|V^{}_{ub}/V^{}_{cb}|$ is naturally smaller than
$|V^{}_{td}/V^{}_{ts}|$ in the heavy quark mass limits. Taking
$c^{}_1 =2$ and $c^{}_2 =1$ for example, we obtain
$|V^{}_{ub}/V^{}_{cb}| \simeq 0.093$ and $|V^{}_{td}/V^{}_{ts}|
\simeq 0.222$ from Eq. (\ref{eq30}), consistent with current data
$|V^{}_{ub}/V^{}_{cb}| \simeq 0.085$ and $|V^{}_{td}/V^{}_{ts}|
\simeq 0.214$ in Eq. (\ref{eq3}). Given $m^{}_t \simeq 172$ GeV and
$m^{}_b \simeq 2.9$ GeV at $M^{}_Z$, one may argue that $m^{}_t \to
\infty$ is a much better limit and thus the relation
$|V^{}_{td}/V^{}_{ts}| \simeq \sqrt{m^{}_d/m^{}_s}$ has a good
chance to be true. In comparison, $|V^{}_{ub}/V^{}_{cb}| \simeq 2
\sqrt{m^{}_u/m^{}_c}$ suffers from much bigger uncertainties
associated with the values of $m^{}_u$ and $m^{}_c$, and even its
coefficient ``$2$" is questionable.

\section{On the Texture Zeros}

In the lack of a quantitatively convincing flavor theory, one has to
make use of possible flavor symmetries or assume possible texture
zeros to reduce the number of free parameters associated with the
fermion mass matrices, so as to achieve some phenomenological
predictions for flavor mixing and CP violation. Note that the
texture zeros of a given fermion mass matrix mean that the
corresponding matrix elements are either exactly vanishing or
sufficiently suppressed as compared with their neighboring
counterparts. There are usually two types of texture zeros:
\begin{itemize}
\item     They may just originate from a proper choice of the flavor basis, and
thus have no definite physical meaning;

\item     They originate as a natural or contrived consequence of an underlying
discrete or continuous flavor symmetry.
\end{itemize}
A typical example of this kind is the famous Fritzsch mass matrices
with six texture zeros \cite{F78},
\footnote{Given a Hermitian or symmetric mass matrix, a pair of
off-diagonal texture zeros have been counted as one zero in the
literature.},
in which three of them come from the basis transformation and
the others arise from either a phenomenological assumption or a
flavor model (e.g., based on the Froggatt-Nielson mechanism
\cite{FN}). Such zeros allow one to establish a few simple and
testable relations between flavor mixing angles and fermion
mass ratios. If such relations are in good agreement with the relevant
experimental data, they may have a good chance to be close to
the truth --- namely, the same or similar relations should be
predicted by a more fundamental flavor model with much fewer free
parameters. Hence a study of possible texture zeros of fermion mass
matrices {\it does} make some sense to get useful hints about flavor
dynamics that is responsible for the generation of fermion masses
and the origin of CP violation.

The original six-zero Fritzsch quark mass matrices was ruled out in
the late 1980's, because it failed in making the smallness of
$V^{}_{cb}$ compatible with the largeness of $m^{}_t$. A
straightforward extension of the Fritzsch ansatz with five or four
texture zeros have been discussed by a number of authors
\cite{FX00}. Given current experimental data on quark flavor mixing
and CP violation, it is found that only the following five five-zero
Hermitian textures of quark mass matrices are still allowed at the
$2\sigma$ level \cite{Raby}:
\begin{eqnarray}
M^{}_{\rm u} = \left(\begin{matrix} 0 & \times & 0 \cr \times &
\times & \times \cr 0 & \times & \times \cr \end{matrix} \right) \;
, \hspace{0.5cm} M^{}_{\rm d} = \left(\begin{matrix} 0 & \times & 0
\cr \times & \times & 0 \cr 0 & 0 & \times \cr
\end{matrix} \right) \; ;
\label{eq31}
\end{eqnarray}
or
\begin{eqnarray}
M^{}_{\rm u} = \left(\begin{matrix} 0 & \times & 0 \cr \times &
\times & 0 \cr 0 & 0 & \times \cr \end{matrix} \right) \; ,
\hspace{0.5cm} M^{}_{\rm d} = \left(\begin{matrix} 0 & \times & 0
\cr \times & \times & \times \cr 0 & \times & \times \cr
\end{matrix} \right) \; ;
\label{eq32}
\end{eqnarray}
or
\begin{eqnarray}
M^{}_{\rm u} = \left(\begin{matrix} 0 & \times & 0 \cr \times & 0 &
\times \cr 0 & \times & \times \cr \end{matrix} \right) \; ,
\hspace{0.5cm} M^{}_{\rm d} = \left(\begin{matrix} 0 & \times & 0
\cr \times & \times & \times \cr 0 & \times & \times \cr
\end{matrix} \right) \; ;
\label{eq33}
\end{eqnarray}
or
\begin{eqnarray}
M^{}_{\rm u} = \left(\begin{matrix} 0 & 0 & \times \cr 0 & \times &
\times \cr \times & \times & \times \cr \end{matrix} \right) \; ,
\hspace{0.5cm} M^{}_{\rm d} = \left(\begin{matrix} 0 & \times & 0
\cr \times & \times & 0 \cr 0 & 0 & \times \cr \end{matrix} \right)
\; ;
\label{eq34}
\end{eqnarray}
or
\begin{eqnarray}
M^{}_{\rm u} = \left(\begin{matrix} 0 & 0 & \times \cr 0 & \times &
0 \cr \times & 0 & \times \cr \end{matrix} \right) \; ,
\hspace{0.5cm} M^{}_{\rm d} = \left(\begin{matrix} 0 & \times & 0
\cr \times & \times & \times \cr 0 & \times & \times \cr
\end{matrix} \right) \; .
\label{eq35}
\end{eqnarray}
In comparison, the Hermitian $M^{}_{\rm u}$ and $M^{}_{\rm d}$ may
also have a parallel structure and contain four texture zeros
\cite{Du}:
\begin{eqnarray}
M^{}_{\rm u} = \left(\begin{matrix} 0 & \times & 0 \cr
\times & \times & \times \cr 0 & \times & \times \cr \end{matrix} \right)
\; , \hspace{0.5cm}
M^{}_{\rm d} = \left(\begin{matrix} 0 & \times & 0 \cr
\times & \times & \times \cr 0 & \times & \times \cr \end{matrix} \right)
\; ,
\label{eq36}
\end{eqnarray}
where only a single zero does not originate from the basis
transformation. But it has been found that a finite (1,1) matrix
element of $M^{}_{\rm u}$ or $M^{}_{\rm d}$ does not significantly
affect the main phenomenological consequences of Eq. (\ref{eq36}),
if its magnitude is naturally small (.e., $\lesssim m^{}_u$ or
$\lesssim m^{}_d$) \cite{Verma}.

The hierarchical structures of four-zero quark mass matrices in Eq.
(\ref{eq36}) can be approximately illustrated as follows
\cite{FX03}:
\begin{eqnarray}
M^{}_{\rm u} & \sim & m^{}_t \left(\begin{matrix} 0 &
\vartheta^3_{\rm u} & 0 \cr \vartheta^3_{\rm u} & ~ \epsilon^2_{\rm
u} ~ & \epsilon^{}_{\rm u} \cr 0 & \epsilon^{}_{\rm u} & 1 \cr
\end{matrix} \right) \; , \nonumber \\
M^{}_{\rm d} & \sim & m^{}_b \left(\begin{matrix} 0 &
\vartheta^3_{\rm d} & 0 \cr \vartheta^3_{\rm d} & ~ \epsilon^2_{\rm
d} ~ & \epsilon^{}_{\rm d} \cr 0 & \epsilon^{}_{\rm d} & 1 \cr
\end{matrix} \right) \; ,
\label{eq37}
\end{eqnarray}
where $\vartheta^{}_{\rm u}$ and $\vartheta^{}_{\rm d}$ essentially
correspond to the definitions in Eq. (\ref{eq14}), and they are
related to $\epsilon^{}_{\rm u}$ and $\epsilon^{}_{\rm d}$ in the
following way:
\begin{eqnarray}
\vartheta^2_{\rm u} & \sim & \epsilon^6_{\rm u} \sim
\frac{m^{}_u}{m^{}_c} \sim 2.2 \times 10^{-3} \; ,
\nonumber \\
\vartheta^2_{\rm d} & \sim & \epsilon^6_{\rm d} \sim
\frac{m^{}_d}{m^{}_s} \sim 4.9 \times 10^{-2} \; .
\label{eq38}
\end{eqnarray}
If the phase difference between $M^{}_{\rm u}$ and $M^{}_{\rm d}$ is
$90^\circ$, then it will be straightforward to obtain $|V^{}_{us}|
\simeq |V^{}_{cd}| \sim \sqrt{\vartheta^2_{\rm u} + \vartheta^2_{\rm
d}}$ and $|V^{}_{cb}| \simeq |V^{}_{ts}| \sim |\epsilon^{}_{\rm u} -
\epsilon^{}_{\rm d}|$. While $|V^{}_{td}/V^{}_{ts}| \sim
\vartheta^{}_{\rm d}$ is apparently expected, $|V^{}_{ub}/V^{}_{cb}|
\sim \vartheta^{}_{\rm u}$ must get modified due to the
non-negligible contribution of ${\cal O}(\vartheta^2_{\rm d}) \sim
{\cal O}(\vartheta^{}_{\rm u})$ from the down-quark sector. Note
that $\epsilon^{}_{\rm u}$ and $\epsilon^{}_{\rm d}$ are not very
small \cite{FX03}, and their partial cancellation results in a small
$|V^{}_{cb}|$ or $|V^{}_{ts}|$.

It is worth pointing out that one may also relax the Hermiticity of
quark mass matrices with a number of texture zeros, such that they
can fit current experimental data very well \cite{XZ2011}. On the
other hand, some texture zeros of quark mass matrices are not
preserved to all orders or at any energy scales in a given flavor
model. If the model is built at a superhigh-energy scale, where a
proper favor symmetry can be used to constrain the structures of
quark mass matrices, one has to take account of the
renormalization-group running effects in order to compare its
phenomenological results with the experimental data at the
electroweak  scale \cite{FX00}.

\section{On the Strong CP Problem}

So far we have discussed weak CP violation based on the CKM matrix
$V$ in the SM. Now let us make a brief comment on the strong CP
problem, because it is closely related to the overall phase of quark
mass matrices and may naturally disappear if one of the six quark
masses vanishes. It is well known that there exists a P- and
T-violating term ${\cal L}^{}_\theta$, which originates from the
instanton solution to the $\rm U(1)^{}_{\rm A}$ problem
\cite{Weinberg}, in the Lagrangian of QCD for strong interactions of
quarks and gluons \cite{QCD}. This CP-violating term can be compared
with the mass term of six quarks, ${\cal L}^{}_{\rm m}$, as follows:
\begin{eqnarray}
{\cal L}^{}_\theta & = & \theta \frac{\alpha^{}_{\rm s}}{8\pi} \
G^a_{\mu \nu} \tilde{G}^{a\mu \nu} \; ,
\nonumber \\
{\cal L}^{}_{\rm m} & = & \overline{\left( \begin{matrix}
u \ & c \ & t \ & d \ & s \ & b \ \end{matrix}\right)^{}_{\rm L}} \ {\cal M}
\left(\begin{matrix} u \cr c \cr t \cr
d \cr s \cr b \end{matrix}\right)^{}_{\rm R} + {\rm h.c.} \; ,
\label{eq39}
\end{eqnarray}
where $\theta$ is a free dimensionless parameter characterizing the
presence of CP violation, $\alpha^{}_{\rm s}$ is the strong
fine-structure constant, $G^a_{\mu \nu}$ (for $a=1,2,\cdots,8$)
represent the $\rm SU(3)^{}_{\rm c}$ gauge fields, $\tilde{G}^{a\mu
\nu} \equiv \epsilon^{\mu\nu\alpha\beta}G^a_{\mu\nu}/2$, and $\cal
M$ stands for the overall $6\times 6$ quark mass matrix. The chiral
transformation of the quark fields $q\to \exp(i\phi^{}_q
\gamma^{}_5) q$ (for $q = u, c, t$ and $d, s, b$) leads to the
changes
\begin{eqnarray}
&& \theta \longrightarrow \theta - 2 \sum_q \phi^{}_q \; ,
\nonumber \\
&& \arg\left(\det {\cal M}\right) \longrightarrow
\arg\left(\det {\cal M}\right) + 2 \sum_q \phi^{}_q \; , ~~
\label{eq40}
\end{eqnarray}
where the change of $\theta$ follows from the chiral anomaly
\cite{Anomaly} in the chiral currents
\begin{eqnarray}
\partial^{}_\mu \left(\overline{q} \gamma^\mu \gamma^{}_5 q \right)
= 2i m^{}_q \overline{q} \gamma^{}_5 q + \frac{\alpha^{}_{\rm
s}}{4\pi} G^a_{\mu\nu} \tilde{G}^{a \mu\nu} \; .
\label{eq41}
\end{eqnarray}
Then the effective CP-violating term in QCD, which is invariant under
the above chiral transformation, turns out to be
\begin{eqnarray}
{\cal L}^{}_{\overline\theta} = \overline{\theta}
\frac{\alpha^{}_{\rm s}}{8\pi} \ G^a_{\mu \nu} \tilde{G}^{a\mu \nu} \; ,
\label{eq42}
\end{eqnarray}
in which $\overline{\theta} = \theta + \arg\left(\det {\cal
M}\right)$ is a sum of both the QCD contribution and the electroweak
contribution \cite{Review}. The latter depends on the phase
structure of the quark mass matrix $\cal M$. Because of
\begin{eqnarray}
\left|\det {\cal M}\right| = m^{}_u m^{}_c m^{}_t m^{}_d m^{}_s
m^{}_b \; ,
\label{eq43}
\end{eqnarray}
the determinant of $\cal M$ becomes vanishing in the $m^{}_u \to 0$
(or $m^{}_d \to 0$) limit. In this case the phase of $\det {\cal M}$
is arbitrary, and thus it can be arranged to cancel out $\theta$
such that $\overline{\theta} \to 0$. Namely, QCD would be a
CP-conserving theory if one of the six quarks were massless. But
current experimental data have definitely ruled out the possibility
of $m^{}_u =0$ or $m^{}_d =0$. Moreover, the experimental upper
limit on the neutron electric dipole moment yields
$\overline{\theta} < 10^{-10}$ \cite{EDM}. The strong CP problem is
therefore a theoretical problem of how to explain why
$\overline{\theta}$ appears but so small \cite{Peccei}.

A comparison between weak and strong CP-violating effects might
make sense, but it is difficult to find out a proper measure for
either of them. The issue involves the reference scale and
flavor parameters which may directly or indirectly determine the
strength of CP violation. To illustrate \cite{Xing2012},
\footnote{We admit that running the heavy quark masses $m^{}_c$,
$m^{}_b$ and $m^{}_t$ down to the QCD scale might not make sense
\cite{Koide}. One may only consider the masses of up and down quarks
\cite{Huang} and then propose ${\rm CP^{}_{strong}} \sim m^{}_u
m^{}_d \sin\overline{\theta}/\Lambda^2_{\rm QCD}$ as an alternative
measure of strong CP violation.}
\begin{eqnarray}
&& {\rm CP^{}_{\rm weak}} \sim \frac{1}{\Lambda^6_{\rm EW}}
\left(m^{}_u - m^{}_c\right) \left(m^{}_c - m^{}_t\right)
\left(m^{}_t - m^{}_u\right) \left(m^{}_d - m^{}_s\right)
\nonumber \\
&& \hspace{1.8cm} \times
\left(m^{}_s - m^{}_b\right) \left(m^{}_b - m^{}_d\right) {\cal
J}^{}_{\rm q} \sim 10^{-13} \; ,
\nonumber \\
&& {\rm CP^{}_{\rm strong}} \sim \frac{1}{\Lambda^6_{\rm QCD}}
m^{}_u m^{}_c m^{}_t m^{}_d m^{}_s m^{}_b \sin\overline{\theta} \sim
10^{4} \sin\overline{\theta} < 10^{-6} \; , ~~~~
\label{eq44}
\end{eqnarray}
where $\Lambda^{}_{\rm EW} \sim 10^2$ GeV, $\Lambda^{}_{\rm QCD}
\sim 0.2$ GeV, and the sine function of $\overline{\theta}$ has been
adopted to take account of the periodicity in its values. So
the effect of weak CP violation would vanish if the masses of any two
quarks in the same (up or down) sector were equal
\footnote{In this special case one of the three mixing angles of $V$
must vanish, leading to ${\cal J}^{}_{\rm q} =0$ too \cite{Mei}.},
and the effect of strong CP violation would vanish if $m^{}_u \to 0$
or $\sin\overline{\theta} \to 0$ held. The remarkable suppression
of CP violation in the SM implies that an interpretation of the
observed matter-antimatter asymmetry of the Universe \cite{PDG}
requests for a new source of CP violation beyond the SM, such as
leptonic CP violation in the decays of heavy Majorana neutrino based
on the seesaw and leptogenesis mechanisms \cite{FY}.

\section{Concluding Remarks}

\begin{figure}[t]
\centerline{\includegraphics[width=6cm]{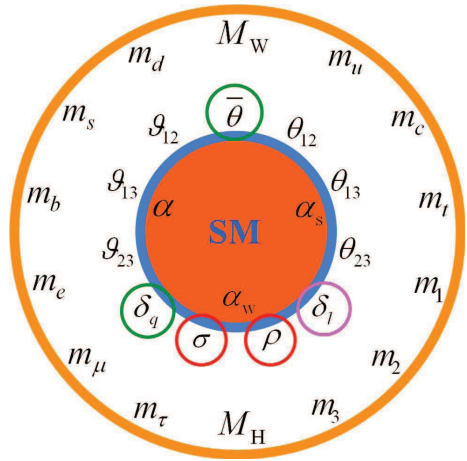}}
\caption{The Fritzsch-Xing ``pizza" plot of 28 parameters associated
with the SM itself and neutrino masses, lepton flavor mixing angles
and CP-violating phases.} \label{fig4}
\end{figure}

Let us make some concluding remarks with the help of the
Fritzsch-Xing ``pizza" plot as shown in Fig. \ref{fig4}. It offers a
summary of 28 free parameters associated with the SM itself and
neutrino masses, lepton flavor mixing angles and CP-violating
phases. Here our focus is on the five parameters of strong and weak
CP violation. In the quark sector, the strong CP-violating phase
$\overline{\theta}$ remains unknown, but the weak CP-violating phase
$\delta^{}_q$ has been determined to a good degree of accuracy. In
the lepton sector, however, none of the CP-violating phases has been
measured. While the Dirac CP-violating phase $\delta^{}_\ell$ can be
determined in the future long-baseline neutrino oscillation
experiments, how to probe or constrain the Majorana CP-violating
phases $\rho$ and $\sigma$ is still an open question.

Perhaps some of the flavor puzzles cannot be resolved unless we
finally find out the fundamental flavor theory. But the latter
cannot be achieved without a lot of phenomenological and
experimental attempts. As Leonardo da Vinci emphasized, ``Although
nature commences with reason and ends in experience, it is necessary
for us to do the opposite. That is, to commence with experience and
from this to proceed to investigate the reason."

Of course, we have learnt a lot about flavor physics from the quark
sector, and are learning much more in the lepton sector. We find
that the flavors of that big pizza in Fig. \ref{fig4} are very
appealing to us.

\section{Acknowledgments}

I am deeply indebted to Harald Fritzsch for inviting me to write
this brief review article and for fruitful collaboration in flavor
physics from which I have benefitted a lot. This work is supported
in part by the National Natural Science Foundation of China under
grant No. 11375207 and the National Key Basic Research Program of 
China under contract No. 2015CB856700.


\end{document}